\documentclass[12pt]{article}
\usepackage[utf8]{inputenc}
\usepackage{amsmath,amssymb}
\usepackage{authblk}
\usepackage{tipa}
\usepackage{listings}
\usepackage{graphicx}
\usepackage [english]{babel}
\usepackage [autostyle, english = american]{csquotes}
\MakeOuterQuote{"}
\usepackage{float}
\usepackage{physics}
\usepackage[super, comma, sort&compress]{natbib}
\usepackage[a4paper, total={7in, 10in}]{geometry}
\usepackage{hyperref}

\title{Qubit vitrification and entanglement criticality on a quantum simulator}

\author[1]{Jeremy C\^{o}t\'{e}}
\author[1]{Stefanos Kourtis\thanks{stefanos.kourtis@usherbrooke.ca}}

\affil[1]{Institut quantique \& D\'{e}partement de physique, Universit\'{e} de Sherbrooke, Sherbrooke, Qu\'{e}bec J1K 2R1, Canada}

\begin{document}

\maketitle

\begin{abstract}
  Many elusive quantum phenomena emerge from a quantum system interacting with its classical environment. Quantum simulators enable us to program this interaction by using measurement operations. Measurements generally remove part of the entanglement built between the qubits in a simulator. While in simple cases entanglement may disappear at a constant rate as we measure qubits one by one, the evolution of entanglement under measurements for a given class of quantum states is generally unknown. We show that consecutive measurements of qubits in a simulator can lead to criticality, separating two phases of entanglement. Using up to 48 qubits, we prepare an entangled superposition of ground states to a classical spin model. Progressively measuring the qubits drives the simulator through an observable vitrification point and into a spin glass phase of entanglement. Our findings suggest coupling to a classical environment may drive critical phenomena in more general quantum states.
\end{abstract}

\section{Introduction}

The Born rule, which states that the outcome of a measurement performed on a quantum state is a random variable whose probability distribution is determined by quantum theory, governs information transfer from a quantum system to its classical environment. While the Born rule is at play constantly all around us since all matter is fundamentally quantum, its effects are only evidenced in bulk due to the astronomical number of measurement events that occur at macroscopic length and time scales. In contrast, a quantum simulator, a programmable array of qubits, is an otherwise isolated quantum system that we can couple to its environment at will with tailor-made measurements of all or some of the qubits. As such, it allows us to study in detail how the quantum characteristics of a system change as we progressively measure its components.

If we were to pick the state of an ideal quantum simulator uniformly at random from the set of all states accessible to it, we would get a \emph{volume-law state}: a state whose \emph{entropy of entanglement} of an extensive subsystem, measured in bits, is proportional to the number of qubits. Measuring a single qubit in such a state removes roughly one bit of entanglement. One therefore expects that the entanglement of a volume-law state should decrease as a simple linear function of the number of measured qubits, yielding a classical unentangled state once we measure all the qubits.

However, this is not always the case: the behaviour of entanglement in a quantum simulator can change dramatically and abruptly as we progressively measure its qubits, exhibiting the phenomenon of \emph{criticality}~\cite{cote_entanglement_2021}. Critical behaviour indicates a transition between two distinct \emph{phases} of entanglement. Previous work revealed entanglement phase transitions in different settings, namely, in random ensembles of quantum circuits in which qubits undergo measurement at a finite rate~\cite{li_measurement-driven_2019, skinner_measurement-induced_2019, ippoliti_entanglement_2021, potter_entanglement_2021, noel_observation_2021, koh_experimental_2022, fisher_random_2022} and in models with topological order~\cite{lavasani_topological_2021, lavasani_monitored_2022, sriram_topology_2022}.

Detecting and characterizing such critical behaviour in experiments is challenging though. First, current quantum processors are faulty, limiting the quantum gates we can reliably implement and the number of error-free measurements we can obtain. Second, measuring the entanglement entropy of an arbitrary quantum state is hard. The straightforward approach requires tomographic reconstruction of the quantum state, which is intractable for quantum systems with many components. Finally, although theoretical models for entanglement phase transitions have been introduced in the context of random circuit ensembles~\cite{li_measurement-driven_2019, skinner_measurement-induced_2019, potter_entanglement_2021, fisher_random_2022}, these are only approximate in the experimentally relevant limit.

Here, we program two entanglement phases and the criticality between them on a quantum simulator of up to 48 superconducting qubits. We do this by implementing an ensemble of quantum circuits~\cite{cote_entanglement_2021} that allow us to reliably generate volume-law states whose entanglement we can deliberately decrease with qubit measurements, experimentally determine the entanglement entropy, and capture the exact dependence of entanglement on measurements using a physical theory~\cite{cote_entanglement_2021} (see ``Measuring entanglement''). The pertinent physical theory is that of spin \emph{vitrification}, i.e., the transition to a spin glass phase~\cite{mezard_nature_1984, schirber_nobel_2021}. We detect the vitrification point, which agrees with spin glass theory. Our work shows measurements alone can trigger entanglement criticality, suggesting a classical environment could induce critical behaviour in more general quantum states.

\section{Results}
\subsection{Theory and model}
\label{sec:TheoryModel}

Our experimental system is an array of superconducting qubits on a quantum simulator. To drive these qubits through an entanglement phase transition, we first execute quantum circuits that prepare highly entangled states. The circuits and theory that follow come from previous theoretical work~\cite{cote_entanglement_2021} on entanglement phase transitions.

\begin{figure}[t]
    \centering
    \includegraphics[width=1.0\textwidth]{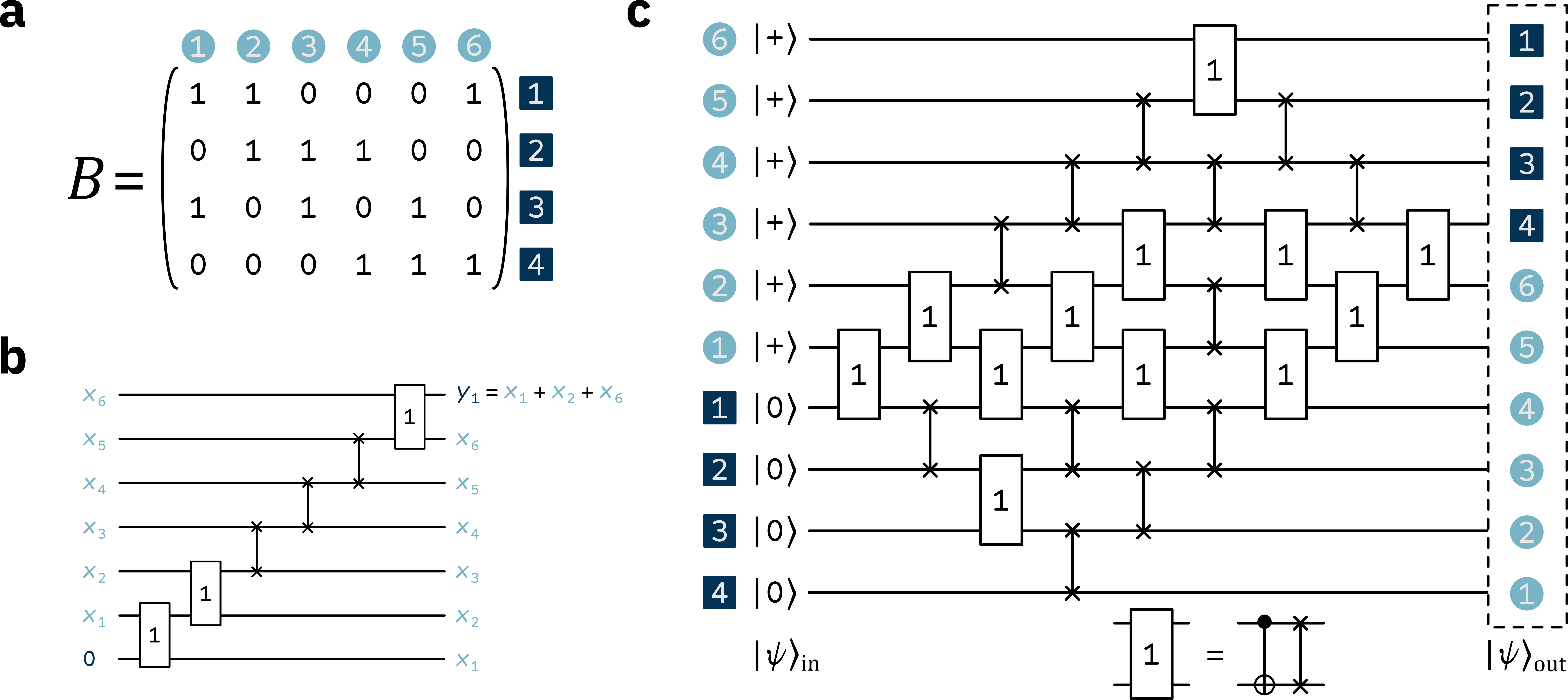}
    \caption{An example of system (Equation~\eqref{eq:MatrixEquation}) with $L = 6$ and $R = 4$. a) The matrix $B$ of Equation~\eqref{eq:MatrixEquation}, where each column represents a variable and each row represents an equation. b) The first linear equation in $B$ compiled as a circuit. The gate labelled ``1'' is a CNOT followed by a SWAP, as shown in the inset of panel c). We use this circuit design to construct the full quantum circuit for $B\vb{x} = \vb{y}$. c) The quantum circuit built from $B$. $B_{ij}=1$ corresponds to the two-qubit gate shown in the inset, whereas $B_{ij}=0$ corresponds to a SWAP gate. The output state $\ket{\psi}_{\mathrm{out}}$ holds all $\vb{y}$ that yield solutions to Equation~\eqref{eq:MatrixEquation} and the corresponding $\vb{x}$ for the matrix $B$. Throughout all panels, the variable labels are in light blue and the parity/equation labels are in dark blue.}
    \label{fig:Model}
\end{figure}

Each of our circuits implements a system of $R$ linear equations on $L$ Boolean variables using $R+L=N$ qubits. (In practice, we use fewer qubits to implement our system on hardware. See ``Circuit optimization'' in the Methods.) We can write the system as the matrix equation
\begin{equation}
    \label{eq:MatrixEquation}
    B \vb{x} = \vb{y} \mod{2} \,,
\end{equation}
where $B$ is an $R \times L$ Boolean matrix whose rows represent equations and columns represent variables, as shown in Fig.~\ref{fig:Model}a. If $B_{ij} = 1$, then equation $i$ involves variable $j$. Otherwise, the entry is zero. By setting the elements of $B$ according to some distribution, we get an ensemble of matrices. Each element of the parity vector $\vb{y} \in \left\{0,1 \right\}^R$ fixes the parity of an equation and each $\vb{x} \in \left\{0,1 \right\}^L$ is a solution to the system for a given $\vb{y}$.

To implement the system of Equation~\eqref{eq:MatrixEquation} using a quantum circuit, we organize the qubits in the simulator in two registers, as sketched in Figs.~\ref{fig:Model}b,c: a ``variable'' register consisting of $L$ variable qubits and a ``parity'' register consisting of $R$ parity qubits. We input variable qubits in the $\ket{+}$ state and parity qubits in the $\ket{0}$ state into a circuit from the ensemble defined above. The initial state is therefore $\ket{\psi}_{\mathrm{in}} = \ket{+}^{\otimes L} \ket{0}^{\otimes R}$. The state $\ket{\psi}_{\mathrm{out}}$ at the output of the circuit is an entangled equal superposition of solutions $\vb{x}$ for each possible $\vb{y}$ (see ``Quantum state and entanglement entropy'' in the Methods for the exact state). For each $\vb{y}$, the set of solutions $\left\{ \vb{x} \right\}$ is unique. Each parity qubit at the output holds the parity of the variables that appear in the corresponding row of $B$ (see Fig.~\ref{fig:Model}b). Because these variable qubits are in a superposition of classical states, so is the parity qubit: it is 0 or 1 depending on the state of the variable qubits. The variable qubits are thus entangled with the parity qubit and contribute one bit of entanglement.

What is the total entanglement between variable and parity qubits? We quantify this with the entanglement entropy $S$, which counts the number of bits of entanglement between two parts of a quantum system. To answer our question above, we need to know how many possible vectors $\vb{y}$ there are in the superposition. For each of the $\mathrm{rank}(B)$ linearly independent rows of $B$, the corresponding component in $\vb{y}$ can be set to zero or one freely. There are then $N_{\vb{y}} = 2^{\mathrm{rank}(B)}$ possible vectors $\vb{y}$ which give solutions to Equation~\eqref{eq:MatrixEquation}. This means the entanglement entropy between the variable and parity qubits is $S(\ket{\psi}_{\mathrm{out}}) \sim \log_2{N_{\vb{y}}} = \mathrm{rank}(B)$.

\subsection{Entanglement phase transition}
\label{sec:PhaseTransition}

Each time we measure a qubit in a generic volume-law state, the system loses one bit of entanglement. This happens with every measurement, reducing the number of superposed configurations until just a single classical state remains.

The states $\ket{\psi}_{\mathrm{out}}$ behave in a manifestly different manner. To prove this, we start by compiling enough linear equations into our circuits such that an output state is a superposition of $N_{\vb{y}} = 2^{\mathrm{rank}(B)} = 2^L$ vectors $\vb{y}$. Then, we calculate the entanglement entropy after measuring a subset $M$ of the parity qubits in our system in the computational basis. We label the measurement outcome $\vb{y}_{\mathrm{out},M}$ and the partially measured state $\ket{\psi}_{\mathrm{out},M}$. The state $\ket{\psi}_{\mathrm{out},M}$ still contains an equal superposition of solutions to Equation~\eqref{eq:MatrixEquation}, but the elements of the parity vector $\vb{y}$ that correspond to the subset $M$ are fixed to the measurement outcome $\vb{y}_{\mathrm{out},M}$. For the same reasoning as in the previous section, there are $N_{\vb{y}_{\mathrm{out},M}} = 2^{\mathrm{rank}(B_M)}$ equally probable measurement outcomes $\vb{y}_{\mathrm{out},M}$, where $B_M$ are the rows of $B$ that correspond to the $M$ parity qubits. (In Fig.~\ref{fig:Model}c, measuring the first three parity qubits would mean $B_M$ is the first three rows of $B$.) Since measuring the output state determines $\vb{y}_{\mathrm{out},M}$, there are $N_{\vb{y}} / N_{\vb{y}_{\mathrm{out},M}} = 2^{L - \mathrm{rank}(B_M)}$ vectors $\vb{y}$ remaining in the state $\ket{\psi}_{\mathrm{out},M}$. The entanglement entropy is then $S(\ket{\psi}_{\mathrm{out}, M}) \sim \log_2{\left(N_{\vb{y}} / N_{\vb{y}_{\mathrm{out},M}} \right)} = L - \mathrm{rank}(B_M)$. Therefore, the evolution of the entanglement entropy is given by $\mathrm{rank}(B_M)$ as a function of $M$. To obtain a size-independent control parameter, we define $\alpha \equiv |M|/L$, the ratio of measured parity qubits to variable qubits.

A unique feature of states $\ket{\psi}_{\mathrm{out}}$ is that we can obtain an exact result for the evolution of the entanglement entropy under measurements through a mapping to a classical spin model~\cite{cote_entanglement_2021}. In this description, $\ket{\psi}_{\mathrm{out}}$ is an entangled superposition of ground states $\vb{x}$ to a spin Hamiltonian with couplings defined by $\vb{y}$ (see Methods). We can then use the characteristics of the spin model both to predict the behaviour of the entanglement and, more importantly, to measure it on a quantum simulator, as we discuss in the next section.

To get concrete predictions for our experiments, we now specify the distribution we sample to populate the matrix $B$ and define the ensemble of states $\ket{\psi}_{\mathrm{out}}$. We pick three distinct variables uniformly at random for each equation in Equation~\eqref{eq:MatrixEquation}, and we ensure there are no repeated equations. With this choice, we get an exact correspondence between our output state and the ground states to an instance of the unfrustrated 3-spin model, both of which are given by the solutions to Equation~\eqref{eq:MatrixEquation}~\cite{franz_ferromagnet_2001, ricci-tersenghi_simplest_2001}. This model exhibits a phase transition at $\alpha_c \approx 0.918$. For $\alpha < \alpha_c$, our system corresponds to a paramagnet in the 3-spin model. We thus get a \emph{paramagnetic} phase of entanglement. Here, $\text{rank}(B_M) = |M| = L\alpha$. This happens because there are few rows in $B_M$, which makes it highly probable that they are linearly independent. The entanglement entropy of the quantum system after measuring $|M|$ parity qubits scales linearly in both $L$ and $\alpha$: $S(\ket{\psi}_{\mathrm{out}, M}) \sim L \left( 1 - \alpha \right)$, i.e., the output state obeys a volume law. For $\alpha > \alpha_c$, the system enters a spin glass phase. The qubits vitrify, turning into an entangled superposition of spin glass ground states. We thus get a \emph{glassy} phase of entanglement. Now, $\text{rank}(B_M) < |M|$ because there are many rows in $B_M$ and linear independence is lost. The entanglement entropy still scales linearly in $L$ but decreases slower than linearly with increasing $\alpha$.

\begin{figure}[H]
    \centering
    \includegraphics[width=0.85\textwidth]{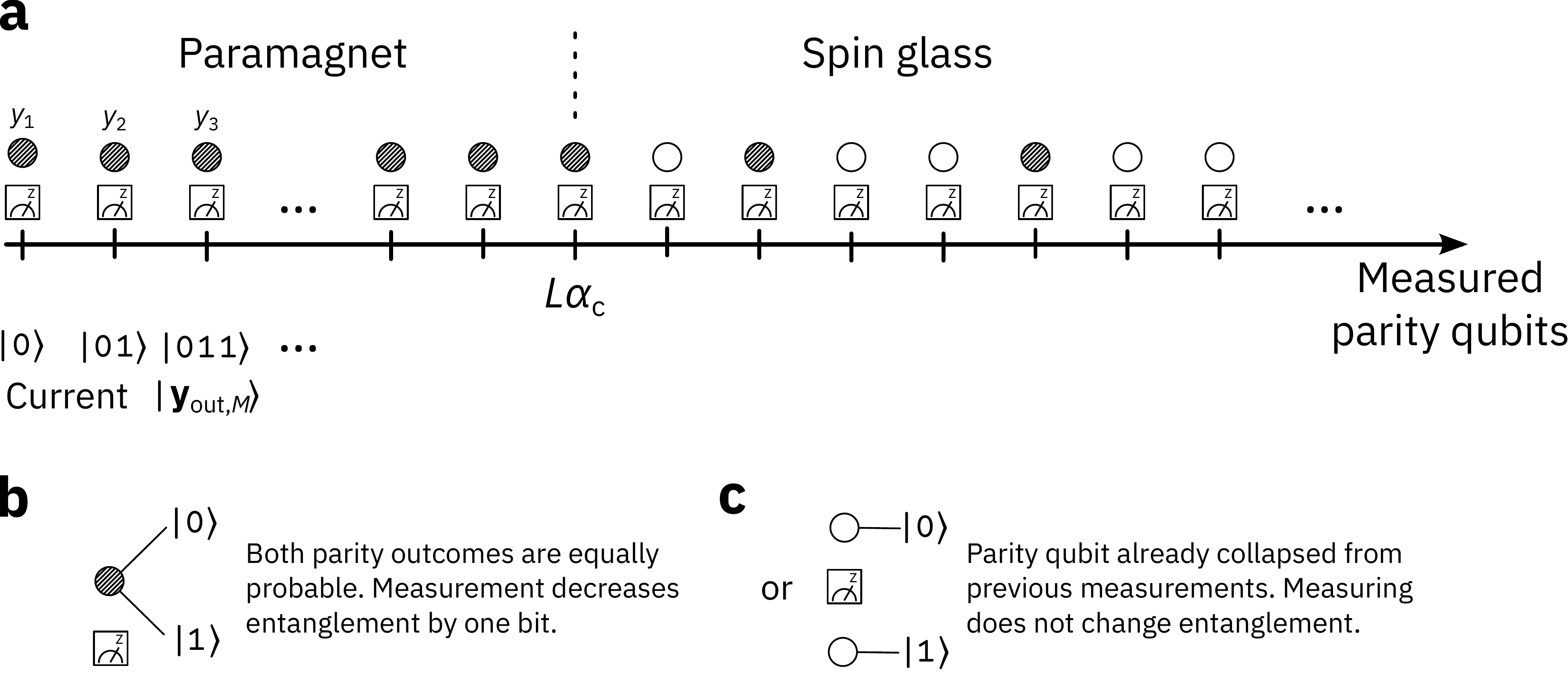}
    \caption{Measurement and collapse in the two entanglement phases. a) Prior to measurement, each parity qubit is either in a superposition as in b), or already collapsed from previous measurements as in c). In the paramagnetic phase, measurement collapses the parity qubit, decreasing the entanglement entropy by one bit. In the spin glass phase, a finite fraction of the parity qubits are already collapsed from previous measurements, so measuring them does not affect the entanglement. An abrupt change between the two behaviours occurs at $|M| = L\alpha_c$.}
    \label{fig:Entanglement}
\end{figure}

We sketch how the measurement of parity qubits collapses the output state in Fig.~\ref{fig:Entanglement}. The two entanglement phases give rise to different behaviours. In the paramagnetic phase, measuring a parity qubit collapses its state to one of two equally probable values (Fig.~\ref{fig:Entanglement}b). This occurs when a parity qubit is independent of the other measured parity qubits, which is the case when $B_M$ has full rank. Measuring the parity qubit halves the number of possible vectors $\vb{y}$ remaining in the superposition, so the system loses one bit of entanglement entropy. In the spin glass phase, there is a finite probability that a parity qubit has a definite value before measurement (Fig.~\ref{fig:Entanglement}c). This occurs when previous measurement outcomes determine the measurement outcome for the next parity qubit, which begins when $\mathrm{rank}(B_M)<|M|$. In this case, measuring the parity qubit does not change the number of possible vectors $\vb{y}$, so the entanglement entropy remains the same.

\subsection{Measuring entanglement}

The exact correspondence between spin glass physics and entanglement established above and in ref.~\cite{cote_entanglement_2021} gives us an efficient way to detect the entanglement phases and criticality on a quantum simulator. In our setup, the entanglement entropy is characterized by the spin glass order parameter~\cite{sherrington_solvable_1975, edwards_theory_1975, parisi_order_1983}
\begin{equation}
    \label{eq:Overlap}
    q(B_M) = \frac{1}{L} \sum_{i = 1}^L \langle (-1)^{x_i} \rangle^2,
\end{equation}
where $\langle \ldots \rangle$ is an average over all the solutions $\vb{x}$ for Equation~\eqref{eq:MatrixEquation} with matrix $B_M$ and a given parity vector $\vb{y}_{\mathrm{out},M}$, and $x_i$ is the $i$-th variable in $\vb{x}$. We derive the identity that links this order parameter to the entanglement entropy in the Methods. Therefore, while in most quantum systems quantifying entanglement is intractable, here we have direct access to the entanglement entropy through the spin glass order parameter, which is efficiently measurable (see Methods).

The order parameter describes the tendency of the solutions $\vb{x}$ to take the same value on each variable. It is zero in the paramagnetic phase, which means a given variable does not have correlations across solutions. The outcome of the measurement of a parity qubit is independent of previous parity measurements in this phase. At $\alpha = \alpha_c$, the variables abruptly become correlated across solutions, and the order parameter jumps to a finite value, eventually saturating to one. This implies solutions of the system are almost identical, differing on only a few variables. The outcome of the measurement of a parity qubit now depends on previous parity measurements.

In the language of physics, each variable can be thought of as one of $L$ spins in a many-body system with $|M|$ interactions. Then, each basis state in $\ket{\psi}_{\mathrm{out},M}$ of the variable qubits represents a spin configuration. In the paramagnetic phase where there are few interactions, these configurations have no correlation, leading to no order ($q = 0$). However, in the spin glass phase where $|M| > L \alpha_c$, the interactions induce correlations across the configurations, leading to spin glass order ($q > 0$).

\begin{figure}[ht]
    \centering
    \includegraphics[width=1.0\textwidth]{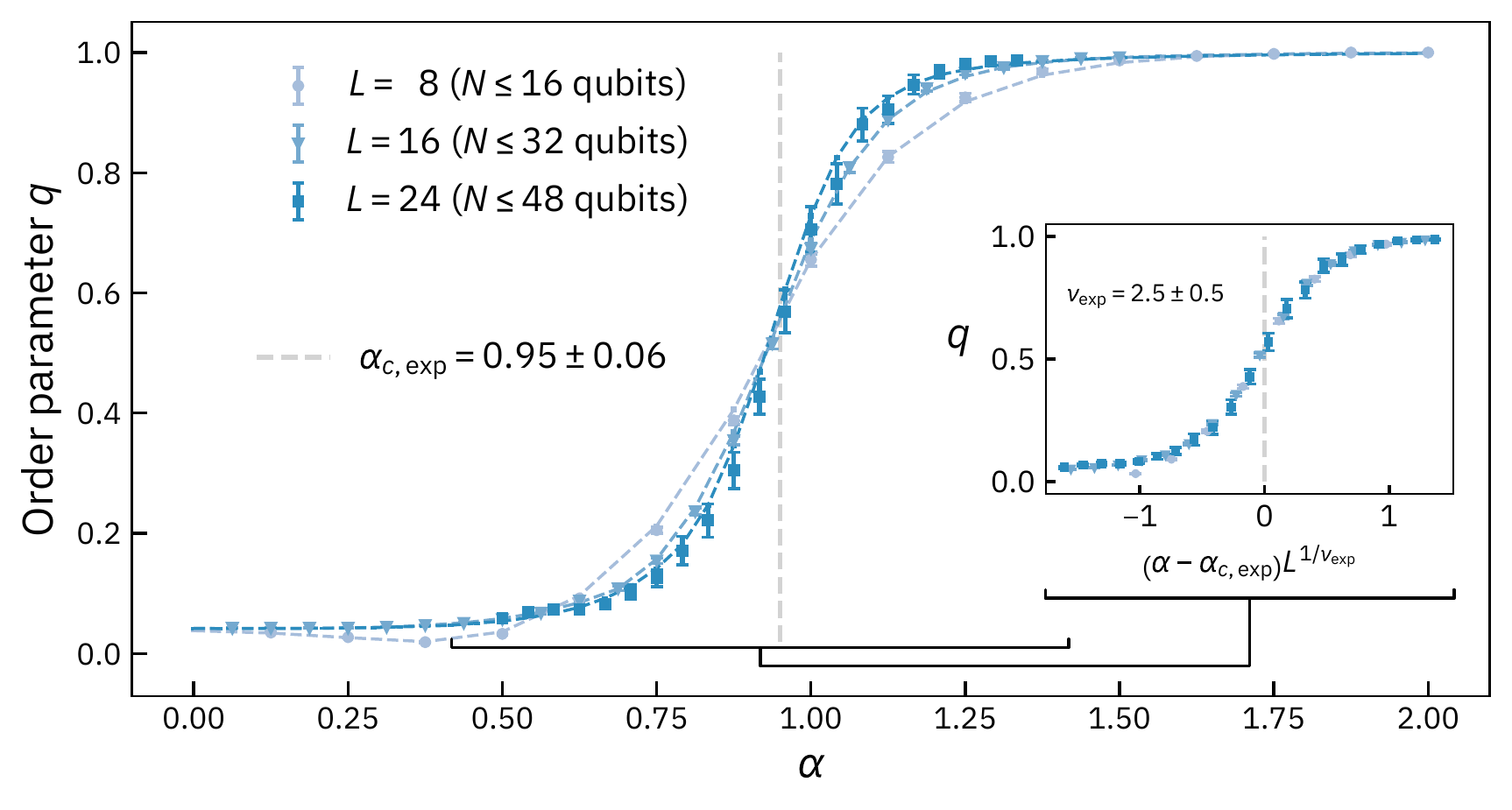}
    \caption{Experimental results for the order parameter $q$ as a function of measurement ratio $\alpha$, using the \textit{ibm\_washington}, \textit{ibmq\_brooklyn} and \textit{ibm\_hanoi} quantum processors~\cite{noauthor_ibm_2021}. Each data point represents an average of the order parameter over $900$ matrices $B_M$, except for $L = 24$, where we average over $50$ matrices. Error bars indicate the standard error of the mean. We study system sizes $L = 8$ (light blue circles), $L = 16$ (medium blue triangles), and $L = 24$ (dark blue squares). Dashed curves provide a reference and connect points from a classical simulation of $q$, where we average over $10,000$ matrices per $\alpha$. The inset shows the collapsed data around $\alpha_{c} \pm 0.5$ with the exponent $\nu_{\mathrm{exp}}$ and critical point $\alpha_{\mathrm{c}, \mathrm{exp}}$ (light gray vertical line), where the curves in the main plot sharpen with increasing system size. For details on data collection, averaging, and finite-size effects, see Methods.}
    \label{fig:experiment}
\end{figure}

Using arrays of up to 48 qubits, our experimental results (Fig.~\ref{fig:experiment}) clearly reveal the two entanglement phases and the transition between them, and are in agreement with theory. The transition at a critical value $\alpha_c$ becomes more abrupt with increasing system size, exactly as spin glass physics dictates. Finite-size scaling (see Methods) of the data using the scaling form $q(\alpha) = f\left(\left(\alpha - \alpha_{c,\mathrm{exp}}  \right) L^{1/\nu_{\mathrm{exp}}} \right)$ yields the experimental values for the critical measurement ratio $\alpha_{c,\mathrm{exp}} = 0.95 \pm 0.06$, which agrees with the theoretical value $\alpha_c \approx 0.918$~\cite{ricci-tersenghi_simplest_2001}, and the critical exponent $\nu_{\mathrm{exp}} = 2.5 \pm 0.5$.

We note that while there are some similarities between the spin glass order we find and those in other work on entanglement criticality~\cite{sang_measurement-protected_2021, li_robust_2021, bao_symmetry_2021}, there are a few differences. First, previous work focuses on states and circuits that respect certain symmetries, which play a role in creating a spin glass phase. In contrast, we do not need to impose symmetries. Second, there is an exact relation between the spin glass order parameter in our work and the entanglement entropy, which is not there in other works. Third, previous work focuses on spin glass order as a steady state property of the system, whereas our system goes into a spin glass state immediately after applying our circuit and measuring. Finally, as our results demonstrate, we can observe this spin glass order on existing quantum hardware.

\section{Discussion}

Since entanglement is a key resource for quantum computation, precise predictions and experimental verification of its possible behaviours in quantum devices under measurement are sought-after. Our findings demonstrate that partial measurements of quantum states can alone give rise to intricate phenomena related to entanglement. Measurements can force qubits to vitrify, and hence realize the celebrated~\cite{schirber_nobel_2021} spin glass phase of matter inside a quantum processor.

The spin glass quantum states implemented here are a subset of stabilizer states, an important class of states for quantum computation. Moreover, we already know that spin glass entanglement criticality is also present in more general classes of states than the ones studied here~\cite{cote_entanglement_2021}. It is interesting to ask whether similar physics applies to entanglement in monitored quantum systems at large, giving rise to different types of nonanalyticity for the entanglement entropy.

\section{Methods}

\subsection{Spin Hamiltonian}

The output of our circuits provides $\vb{x}$ and $\vb{y}$ from Equation~\eqref{eq:MatrixEquation}, which we can relate to the ground states and couplings of a $p$-spin model~\cite{mezard_two_2003} (with $p = 3$). The model consists of $L$ spins, with $R$ interactions encoded by the matrix $B$ (see the example in Fig.~\ref{fig:Model}a). The indices of the nonzero elements in each row $a \in B$ correspond to the spins which are part of an interaction. The Hamiltonian is:
\begin{equation}
    \label{eq:Hamiltonian}
    H(B, \boldsymbol{\sigma}, \vb{J}) = \frac{1}{2} \sum_{a \in B} \left( 1 - J_a \sigma_{a_1} \sigma_{a_2} \sigma_{a_3} \right),
\end{equation}
where $a_1, a_2, a_3$ refer to three distinct spins (the indices of the nonzero elements in $a$), $J_a$ are the couplings of the interaction vector $\vb{J} \in \left\{\pm 1 \right\}^R$, and the spins $\sigma_i$ form the spin vector $\boldsymbol{\sigma} \in \left\{ \pm 1 \right\}^L$. The ground-state energy for this Hamiltonian is zero, which corresponds to $J_a \sigma_{a_1} \sigma_{a_2} \sigma_{a_3} = 1$ for all $a$.

Using the mapping $J_a = (-1)^{y_a}$ and $\sigma_{i} = (-1)^{x_{i}}$, we see that Equation~\eqref{eq:Hamiltonian} is zero whenever $y_a = x_{a_1} + x_{a_2} + x_{a_3} \mod 2$ for all $a$, which is Equation~\eqref{eq:MatrixEquation}. This lets us express the number of ground states $\mathcal{N}_{\mathrm{GS}}$ to Equation~\eqref{eq:Hamiltonian} in terms of $\vb{y}$ and $B$. Because each of the $N_{\vb{y}} = 2^{\text{rank}(B)}$ vectors $\vb{y}$ has $\mathcal{N}_{\mathrm{GS}}$ ground states (out of a possible $2^L$), we find $\mathcal{N}_{\mathrm{GS}} = 2^{L - \text{rank}(B)}$. The ground state entropy is $S_{\mathrm{GS}} (B) \equiv \log \mathcal{N}_{\mathrm{GS}} = \left[ L - \text{rank}(B) \right] \log 2$ (we take the natural logarithm).

\subsection{Quantum state and entanglement entropy}
After applying the circuit given by $B$ (Fig.~\ref{fig:Model}c) to our input state $\ket{\psi}_{\mathrm{in}}$, we get the following state (see ref.~\cite{cote_entanglement_2021} for more details):

\begin{equation}
    \label{eq:OutputState}
    \ket{\psi}_{\mathrm{out}} = \frac{1}{\sqrt{N_{\vb{y}}}}  \sum_{ \vb{y}} \ket{\vb{y}} \ket{\{ \vb{x} \, : \,  B\vb{x} = \vb{y} \}} \,.
\end{equation}

This is a superposition of solutions $\left\{ \vb{x} \right\}$ for each of the $N_{\vb{y}}$ possible $\vb{y}$. We then measure the state of the first $|M|$ parity qubits to be $\vb{y}_{\mathrm{out}, M}$. The resulting state is

\begin{equation}
    \label{eq:MeasuredState}
    \ket{\psi}_{\mathrm{out},M} = \frac{1}{\sqrt{N_{\vb{y}} / N_{\vb{y}_{\mathrm{out},M}}}}  \sum_{ \{ \vb{y} \,:\, \vb{y}_{|M|} = \vb{y}_{\mathrm{out},M} \} } \ket{\vb{y}} \ket{\{ \vb{x} \, : \,  B\vb{x} = \vb{y} \}} \,,
\end{equation}
where the first $|M|$ components of $\vb{y}$ are $\vb{y}_{|M|} = \vb{y}_{\mathrm{out},M}$. The state is still an equal superposition of solutions for each $\vb{y}$, but now there are only $N_{\vb{y}} /  N_{\vb{y}_{\mathrm{out}, M}}$ terms in the sum, with $N_{\vb{y}_{\mathrm{out}, M}} = 2^{\mathrm{rank}(B_M)}$. The coefficient $\lambda_{\vb{y}} = \sqrt{N_{\vb{y}_{\mathrm{out}, M}} / N_{\vb{y}}}$ determines the entanglement entropy for such a state:

\begin{equation}
    \label{eq:MeasuredEntanglement}
    S(\ket{\psi}_{\mathrm{out},M}) \equiv - \sum_{\{ \vb{y} \,:\, \vb{y}_{|M|} = \vb{y}_{\mathrm{out},M} \}} \lambda_{\vb{y}}^2 \log{\left( \lambda_{\vb{y}} ^2 \right)} = \Big[ \text{rank}(B)-\text{rank}(B_M) \Big] \log 2.
\end{equation}

The entanglement entropy coincides with $S_{\mathrm{GS}}(B_M)$---the ground state entropy of the classical spin model---when we choose our initial $B$ to satisfy $\text{rank}(B) = L$. In the limit of large system sizes, the expression for the averaged entropy density~\cite{monasson_course_2007} is:

\begin{equation}
    \label{eq:MeasuredEntanglement2}
    \lim_{L \rightarrow \infty} \frac{\langle S(\ket{\psi}_{\mathrm{out},M}) \rangle}{L} = \Big[ \left( 1 - q(\alpha) \right) \left( 1 - \log{\left(1 - q(\alpha) \right)} \right) - \alpha \left( 1 - q^3(\alpha) \right) \Big] \log 2,
\end{equation}
where $q(\alpha)$ is the spin glass order parameter after performing the ensemble average using Equation~\eqref{eq:Overlap}. Equation~\eqref{eq:MeasuredEntanglement2} establishes the exact correspondence between the entanglement entropy and the spin glass order parameter.

\subsection{Quantum hardware}

We used the \textit{ibm\_washington}, \textit{ibmq\_brooklyn} and \textit{ibm\_hanoi} quantum processors~\cite{noauthor_ibm_2021} for our experiments. We set the repetition delay to $0.00025 \mathrm{s}$. We chose connected lines of qubits to take advantage of the one-dimensional structure of our circuits, while also having low measurement readout error and CNOT error rates at the time of job submission.

\subsection{Circuit optimization}

Because errors dominate the output in current quantum processors, we optimize our circuits to use as few gates as possible. As the CNOT is the native two-qubit gate on the IBM Q processors, we use the CNOT count $N_{\mathrm{CNOT}}$ as our metric (we ignore the $L$ single-qubit Hadamard gates we always need). We build our circuits using the matrix $B_M$ since it generates the solutions we use in Equation~\eqref{eq:Overlap} and requires fewer gates to implement than $B$. We then decompose SWAP gates in our circuits as
\begin{equation}
    \label{eq:SWAP}
    \text{SWAP}(i,j) = \text{CNOT}(i,j) \times \text{CNOT}(j,i) \times \text{CNOT}(i,j),
\end{equation}
where $i$ and $j$ are the qubits participating in the gate and CNOT$(i,j)$ means qubit $i$ controls the target qubit $j$. For the ``1'' gate in Fig.~\ref{fig:Model}, using Equation~\eqref{eq:SWAP} reveals two consecutive CNOT gates with the same control and target, which we remove because they have no overall effect. As such, a one in the matrix requires two CNOTs while a zero requires three.

How many qubits and CNOT gates do we need to build circuits such as in Fig~\ref{fig:Model}c using $B_M$? There are $|M| = L\alpha$ linear equations, so we require $N = L + |M| = L (1 + \alpha)$ qubits. To calculate $N_{\mathrm{CNOT}}$, note that each row of $B_M$ contains $p$ ones and $L-p$ zeros. There are then $2p + 3(L-p)$ CNOTs per row of $B_M$, where $p=3$ for our model. Summing the CNOT count over all rows, we find $N_{\mathrm{CNOT}}(B_M) = |M| \left[ 2p + 3(L-p) \right] = 3L \left(L - 1\right) \alpha$.

By transforming $B_M$ using matrix row operations, we can reduce $N$ and $N_{\mathrm{CNOT}}$. We begin by putting $B_M$ into row echelon form. Then for each row of the matrix (starting from the second), we find the index of the leading one, and add this row to the rows above it which have a zero at that index. These row additions generate more ones in the matrix, which we prefer because they require less gates than zeros to implement. We call the resulting matrix $B_{M'}$. Note that solutions to Equation~\eqref{eq:MatrixEquation} for a given parity vector remain unchanged under row operations.

For example, consider the following matrix:
\begin{equation}
    \label{eq:example}
  B_M =  \begin{pmatrix}
        1 & 0 & 1 & 0 & 0 & 1 \\
        0 & 1 & 0 & 1 & 0 & 1 \\
        0 & 1 & 1 & 1 & 0 & 0 \\
        0 & 1 & 1 & 0 & 1 & 0 \\
        1 & 1 & 0 & 0 & 0 & 1 
    \end{pmatrix}.
\end{equation}
Applying the operations described gives
\begin{equation}
    \label{eq:Equivalent}
  B_{M'} =   \begin{pmatrix}
        1 & 1 & 1 & 1 & 1 & 0 \\
          & 1 & 1 & 1 & 1 & 0 \\
          &   & 1 & 1 & 1 & 1 \\
          &   &   & 1 & 1 & 0 \\
          &   &   &   & 1 & 0 
    \end{pmatrix},
\end{equation}
where the omitted entries are zeros.

The form of $B_{M'}$ helps us save qubits and gates. First, the matrix has size $|M'| \times L$, with $|M'| = \mathrm{rank}(B_M)$. The resulting circuit requires $N = L + \mathrm{rank}(B_M) \leq L + |M|$ qubits, fewer than the circuits built from $B_M$ when $ \mathrm{rank}(B_M) < |M|$. Second, notice that in Fig~\ref{fig:Model}c, there are gates for each entry of the matrix. By interspersing the parity and variable qubits instead of separating them, we can avoid including gates for the zeros to the left of the leading ones in $B_{M'}$.

Each gate in the primitive circuit (Fig.~\ref{fig:Model}b) exchanges the positions of the qubits it acts upon. The result is that a parity qubit exchanges positions with every variable qubit. However, only ``1'' gates contribute to the parity we want to measure. Therefore, once a parity qubit encounters all the ``1'' gates for its linear equation, we measure it in that position. We take advantage of this by altering the primitive circuit: we start the parity qubit at the top, reverse the gate sequence, and invert the control and target of the CNOTs in each ``1'' gate. Then, the locations of the leading ones in $B_{M'}$ provide the end positions for measuring the parity qubits. For example, the parity qubit for the first row in Equation~\eqref{eq:Equivalent} exchanges positions with all variable qubits before we measure it. The parity qubit for the second row exchanges positions with variable qubits 6, 5, 4, 3, and 2 before measuring, and so on.

We provide an upper bound for $N_{\mathrm{CNOT}}(B_{M'})$. We count the entries in $B_{M'}$ to the right of (and including) the main diagonal. We assume the leading ones are all part of the main diagonal. With this assumption, the number of entries to the right of (and including) the leading ones in a $P \times Q$ row echelon matrix is:

\begin{equation}
    U(P,Q) = \sum_{i = 1}^P \left( Q - [i-1] \right) = P \left( Q -  \frac{1}{2} \left( P - 1 \right) \right).
    \label{eq:Upper}
\end{equation}

In our case, $P = \mathrm{rank}(B_M)$ and $Q = L$. There is at least a one per row (else the row would not be a part of $B_{M'}$). Since zeros contribute more to $N_{\mathrm{CNOT}}$, we take the worst-case scenario where all other entries are zero. This implies there are $P$ ones in the matrix, so there are $Z = U(P,Q) - P$ zeros. Using these results, we get the upper bound:

\begin{equation}
    \label{eq:GateCount}
    N_{\text{CNOT}}(B_{M'}) \leq 3Z + 2P = \frac{3}{2} \mathrm{rank}(B_M) \left[ 2L - \mathrm{rank}(B_M) + \frac{1}{3} \right] \leq \frac{1}{2} L\left( 3L + 1 \right),
\end{equation}
where we get the final inequality by maximizing the previous expression with respect to $\mathrm{rank}(B_M)$.

Finally, rather than putting a variable qubit in its initial superposition $\ket{+} = H \ket{0}$ as an input to the circuit, we apply the Hadamard gate $H$ only when the corresponding variable first participates in a linear equation. (For example, in Equation~\eqref{eq:Equivalent} variable 6 is first part of an equation in row 3.) Doing so reduces errors from trying to maintain superpositions for too long in current quantum processors. It also simplifies any SWAP gate involving a variable qubit in the state $\ket{0}$. If we have qubits $i$ and $j$ with the latter in the state $\ket{0}$, Equation~\eqref{eq:SWAP} reduces to
\begin{equation}
    \label{eq:reducedSWAP}
    \text{SWAP}(i,j)|_{\text{$j$ in $\ket{0}$}} = \text{CNOT}(i,j) \times \text{CNOT}(j,i).
\end{equation}

Our circuit optimization provides a significant savings compared to the circuits built from $B_M$, which require $N_{\mathrm{CNOT}}(B_M) = 3L(L-1) \alpha$ gates and $N = L \left(1 + \alpha \right)$ qubits. In practice, our largest experiments ($L = 24$ and $\alpha \gtrsim 1$) required an average of $N_{\mathrm{CNOT}}(B_{M'}) \approx 600$ gates, which is much less than the $N_{\mathrm{CNOT}}(B_M) \gtrsim 1600$ gates we would need if we used the $B_M$ matrices instead.

\subsection{Error mitigation and shot count}

\begin{figure}[H]
    \centering
    \includegraphics[width=0.8\textwidth]{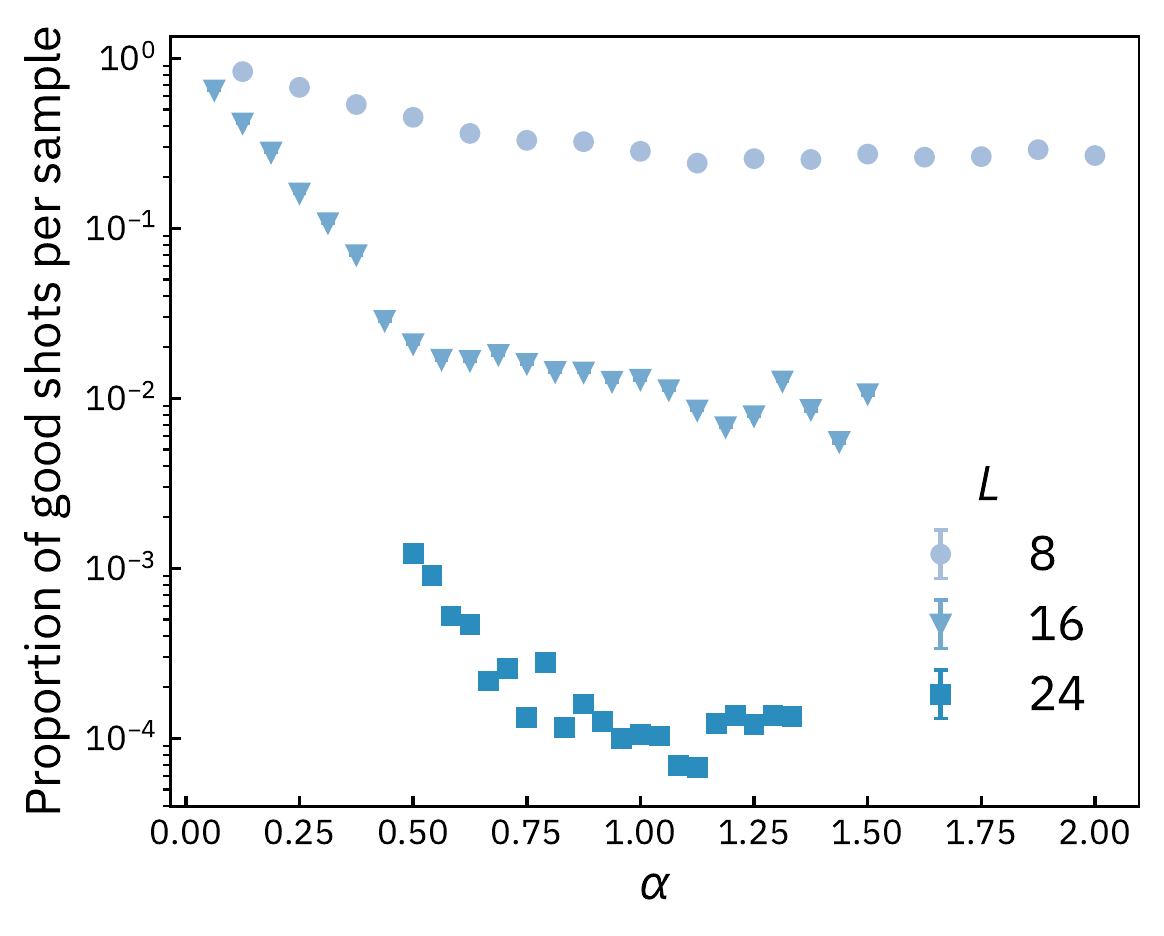}
    \caption{The proportion of shots which satisfy $B_{M'}\vb{x} = \vb{y}_{\mathrm{out},M'}$ for $|M| = L \alpha$ from our experiments on the IBM Q processors. Note the logarithmic vertical scale. We average over $900$ samples for each data point, except for $L = 24$, where we average over $50$ samples. Error bars indicate the standard error of the mean and are mostly smaller than the markers. We study system sizes $L = 8$ (light blue circles), $L = 16$ (medium blue triangles), and $L = 24$ (dark blue squares).}
    \label{fig:goodShotsProportion}
\end{figure}

In our experiments, we only keep measurement results (shots) $\vb{x}$ and $\vb{y}_{\mathrm{out},M'}$ which satisfy $B_{M'}\vb{x} = \vb{y}_{\mathrm{out},M'}$. This provides significant error mitigation (Fig.~\ref{fig:goodShotsProportion}) as we increase the number of qubits. For $L = \left\{ 8, \, 16, \, 24 \right\}$, we took $\left\{10000, \, 25000, \, 750000 \right\}$ shots per sample (see next section) to obtain our data.

\subsection{Data collection}

\begin{enumerate}
    \item For the desired number of matrix samples:
    
    \begin{enumerate}
        \item Generate a random matrix $B$ as described in the ``Entanglement phase transition'' section with $L$ columns and $L\alpha_{\mathrm{max}}$ rows. Each $B$ is a sample and provides data for $\alpha \in \left(0, \alpha_{\mathrm{max}} \right]$.
        
        \item For each $\alpha \in \left(0, \alpha_{\mathrm{max}} \right]$:
        
        \begin{enumerate}
            \item Take the submatrix $B_M$, consisting of the first $|M| = L \alpha$ rows of $B$.
            
            \item Put $B_M$ into row echelon form and perform row operations as described in the ``Circuit optimization'' section. The result is $B_{M'}$.
            
            \item Build the circuit corresponding to the matrix $B_{M'}$ using the techniques described in the ``Circuit optimization'' section.
        
            \item Execute the circuit on the quantum processor a sufficient number of times. Here, sufficient means measuring several pairs $(\vb{x},\vb{y}_{\mathrm{out},M'})$ that pass the test in the following step. We always had at least 18 pairs.
            
            \item Test measurements $(\vb{x},\vb{y}_{\mathrm{out},M'})$ by verifying if $B_{M'} \vb{x} = \vb{y}_{\mathrm{out},M'}$.
            
            \item Save the variable and parity vectors $\vb{x}$ and $\vb{y}_{\mathrm{out},M'}$ that pass the test.
        \end{enumerate}
    
    \end{enumerate}
\end{enumerate}

\subsection{Calculating the order parameter}
\label{sec:Overlap}

\begin{enumerate}

    \item For each $\alpha \in \left(0, \alpha_{\mathrm{max}} \right]$:
    
    \begin{enumerate}
        \item For each matrix $B$ from the previous section:
        
        \begin{enumerate}
            \item Compute $B_{M'}$ as in the previous section using $B_M$ with $|M| = L \alpha$.
            
            \item For each saved parity vector $\vb{y}_{\mathrm{out},M'}$ associated to $B_{M'}$:
            
            \begin{enumerate}
                \item Fix a reference solution $\vb{z}$ that maps solutions from the parity vector $\vb{y}_{\mathrm{out},M'}$ to the parity vector $\vb{0}$. We chose our reference to be the solution to $B_{M'} \vb{z} = \vb{y}_{\mathrm{out},M'}$ whose binary form represents the smallest integer. Note that finding a reference is efficient.
                
                \item Map the saved solutions $\vb{x}$ associated with $\vb{y}_{\mathrm{out},M'}$ to $\vb{x}' = \vb{x} + \vb{z}$. Now, $B_{M'} \vb{x}' = \vb{0}$. Remove any duplicates. Call this set $X = \left\{ \vb{x}' \right\}$.
            \end{enumerate}
            
            \item Compute $q(B_{M'})$ in Equation~\eqref{eq:Overlap} by uniformly sampling $\min \left(24, |X| \right)$ solutions from $X$, where we determined the number 24 yields a reasonable compromise between accuracy and quantum runtime.
            
        \end{enumerate}
        
        \item Compute $q(\alpha)$ by averaging over $q(B_{M'})$ for all $B_{M'}$.
    \end{enumerate}
    
\end{enumerate}

Following the procedure for each $L$ produces the curves in Fig.~\ref{fig:experiment}. To calculate the order parameter in step iii), we want as many solutions $\vb{x}'$ as possible, but a finite number works, making the order parameter efficient to measure. For the classical simulation of $q$ (the dashed lines in Fig.~\ref{fig:experiment}), we uniformly sample $\min \left(24, \mathcal{N}_{\mathrm{GS}} \right)$ solutions to the equation $B_M\vb{x} = \vb{0}$, where $\mathcal{N}_{\mathrm{GS}} = 2^{L - \mathrm{rank}(B_M)}$ is the total number of solutions. We did this by sampling random linear combinations of the basis vectors forming the null space of $B_M$. This is also efficient.

We note that the small size of the sample $X$ leads to appreciable artefacts in Fig.~\ref{fig:experiment}, such as a deviation of the order parameter from the expected value of zero at small $\alpha$. Concomitantly, we notice the onset of finite-size effects at values of $L$ and $\alpha$ for which $\min \left(24, \mathcal{N}_{\mathrm{GS}} \right) \approx \mathcal{N}_{\mathrm{GS}}$. The dip of $q$ at small $\alpha$ for $L=8$ is such an effect. We nevertheless notice that, for all $L$ and $\alpha$, theory and experiment match well in Fig.~\ref{fig:experiment}, since these effects are present in both.

\subsection{Finite-size scaling}

The objective of finite-size scaling is to take the data in Fig.~\ref{fig:experiment} and try to collapse it onto a common curve by finding suitable critical parameters. We follow the technique of ref.~\cite{kawashima_critical_1993} and our previous work~\cite{cote_entanglement_2021}. In particular, we use the scaling form:

\begin{equation}
    \label{eq:scalingForm}
    q = f\left(\left(\alpha - \alpha_{c,\mathrm{exp}}  \right) L^{1/\nu_{\mathrm{exp}}} \right),
\end{equation}
and we minimize a cost function with the data and its associated error as input to find the critical parameters $\alpha_{c,\mathrm{exp}}$ and $\nu_{\mathrm{exp}}$.

We store our experimental data as a triple $\left( \alpha, q(\alpha), e(\alpha) \right)$, where $e(\alpha)$ is the standard error of the mean for the data point. The standard error of the mean for $n$ samples is:
\begin{equation}
    \label{eq:StandardError}
    e(\alpha) = \sqrt{\sum_{i=1}^n \frac{\left[q_i(\alpha) - \Bar{q}(\alpha) \right]^2}{n(n-1)}},
\end{equation}
where $q_i(\alpha)$ is the order parameter for a given matrix $B$ and $\alpha$, and $\Bar{q}(\alpha)$ is the mean of $q_i(\alpha)$ over $i$.

We then transform the triple according to the scaling form:
\begin{equation}
    \label{eq:TransformData}
    \left( t_i, g_i, e_i \right) = \left( \left[ \alpha - \alpha_{c, \mathrm{exp}} \right] L^{1/\nu_{\mathrm{exp}}}, q(\alpha), e(\alpha) \right).
\end{equation}
We sort these triples by their $t$-values and then compute the cost function:
\begin{equation}
    \label{eq:CostFunction}
    C(\alpha_{c, \mathrm{exp}}, \nu_{\mathrm{exp}}) = \frac{1}{T-2} \sum_{i = 2}^{T-1} w\left( t_i, g_i, e_i \vert t_{i-1}, g_{i-1}, e_{i-1}, t_{i+1}, g_{i+1}, e_{i+1}  \right),
\end{equation}
with $T$ being the number of data points. The quantity in the summation is:
\begin{equation}
    \label{eq:w}
    w\left( t_i, g_i, e_i \vert t_{i-1}, g_{i-1}, e_{i-1}, t_{i+1}, g_{i+1}, e_{i+1}  \right) = \left( \frac{g_i - \Bar{g}}{\Delta \left(g_i - \Bar{g} \right)} \right)^2,
\end{equation}

\begin{equation}
    \label{eq:yBar}
    \Bar{g} = \frac{\left(t_{i+1} - t_i \right) g_{i-1} - \left(t_{i-1} - t_i \right) g_{i+1}}{\left(t_{i+1} - t_{i-1} \right)},
\end{equation}

\begin{equation}
    \label{eq:Uncertainty}
    \left[ \Delta \left(g_i - \Bar{g} \right) \right]^2 = e_i^2 + \left( \frac{t_{i+1} - t_{i}}{t_{i+1} - t_{i-1}} \right)^2 e_{i-1}^2 + \left( \frac{t_{i-1} - t_{i}}{t_{i+1} - t_{i-1}} \right)^2 e_{i+1}^2.
\end{equation}

The cost function $C(\alpha_{c, \mathrm{exp}}, \nu_{\mathrm{exp}})$ measures, for each index $i$, the squared deviation of the point $\left(t_i, g_i \right)$ from the linear interpolation $\Bar{g}$ between the points $\left(t_{i-1}, g_{i-1} \right)$ and $\left(t_{i+1}, g_{i+1}\right)$ on either side of the sorted sequence. We exclude the first and last points in the sequence since they have no neighbouring points to the left or right, respectively. The uncertainty (Equation~\eqref{eq:Uncertainty}) is a weighted sum of the squared error of the current point $\left(t_{i}, g_{i} \right)$ and the squared error from the linear interpolation (Equation~\eqref{eq:yBar}). We skip over any three identical $t$-values in a row in Equation~\eqref{eq:CostFunction} because of the division by zero in Equations~\eqref{eq:yBar} and~\eqref{eq:Uncertainty} (though this only happens for isolated values of $\alpha_{c, \mathrm{exp}}$). When this happens, we reduce the denominator of the fraction in front of Equation~\eqref{eq:CostFunction} by the number of skips.

\begin{figure}[ht]
    \centering
    \includegraphics[width=1.0\textwidth]{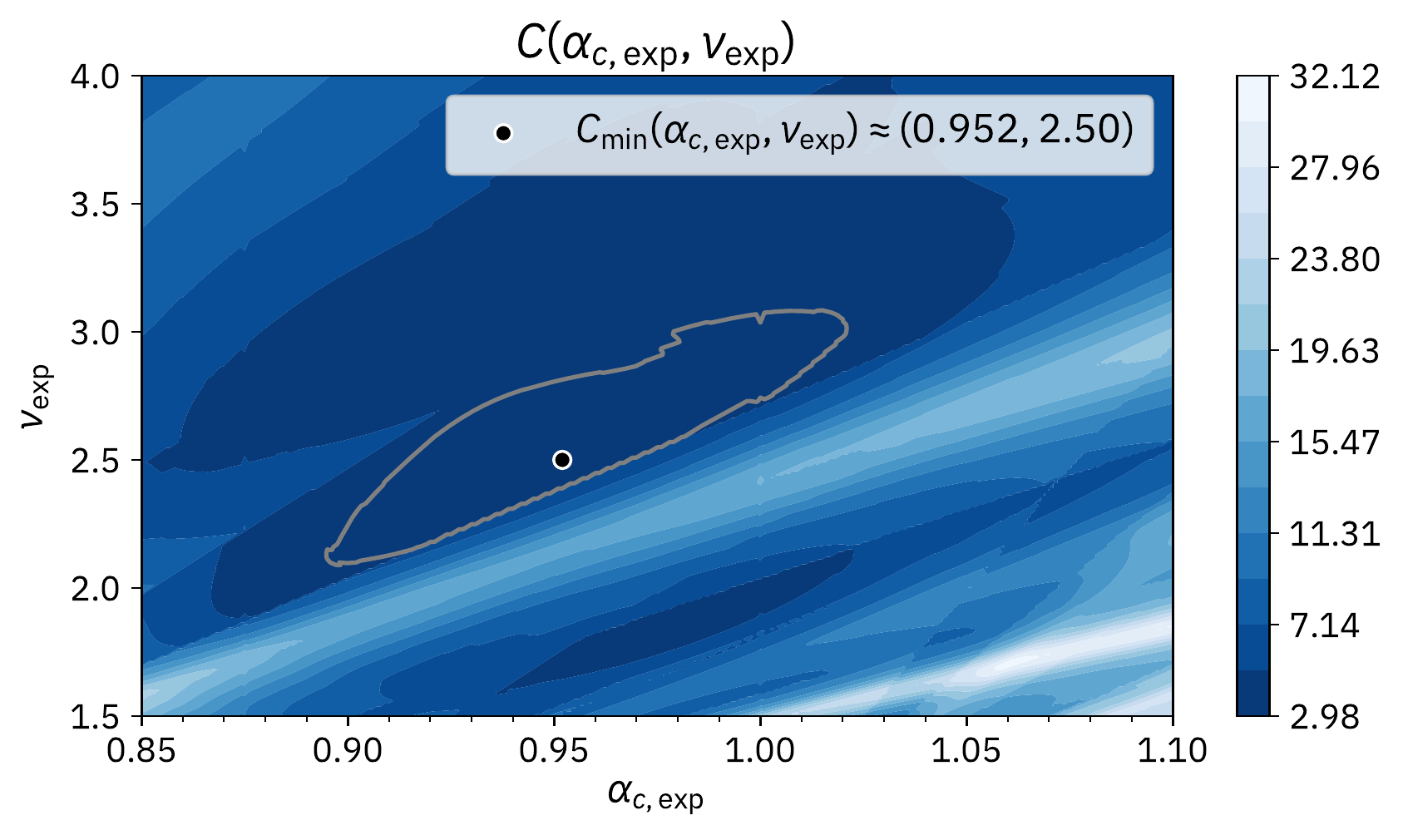}
    \caption{The cost function landscape over a grid of values of $\alpha_{c, \mathrm{exp}}$ and $\nu_{\mathrm{exp}}$. The black dot indicates the minimum of the cost function, located at the optimal $\alpha_{c, \mathrm{exp}}$ and $\nu_{\mathrm{exp}}$ in the legend (to the resolution of the grid). The grey contour marks the region of uncertainty, given by $\left(1 + r\right)C_{\text{min}}(\alpha_{c, \mathrm{exp}}, \nu_{\mathrm{exp}})$, with $r = 0.25$. To compute $C(\alpha_{c, \mathrm{exp}}, \nu_{\mathrm{exp}})$, we used the range $\alpha_c \pm 0.5$ for the data in Equation~\eqref{eq:CostFunction}.}
    \label{fig:Landscape}
\end{figure}

We plot the cost function over a grid of values near the critical parameters from the literature (Fig.~\ref{fig:Landscape}). This allows us to visualize both the minimum and the uncertainty around it.
The collapse for finite-size scaling works best when there are finite-size effects, so we restricted our data for the cost function to the region $\alpha_c \pm 0.5$ (the black connector linking the main plot with the inset in Fig.~\ref{fig:experiment}).

We chose our grid for the critical parameters to be $\alpha_{c, \mathrm{exp}} \in \left[0.85, 1.10\right]$, with a step size of $0.001$, and $\nu_{\mathrm{exp}} \in \left[1.5, 4.0 \right]$, with a step size of $0.01$. We chose a finer step size for $\alpha_{c, \mathrm{exp}}$ because we know the critical threshold. We used a larger step size for $\nu_{\mathrm{exp}}$ because there is less precision in the literature for $\nu$.

To estimate our uncertainty, we plot a contour at the level $\left(1 + r\right) C_{\text{min}}$, where $r$ is the size of the maximum deviation we allow in the minimum value. We chose $r = 0.25$, which means we remain uncertain about the minimum for values that are up to $25\%$ larger. Changing $r$ will grow or shrink the contour. We note in Fig.~\ref{fig:Landscape} that the cost function's minimum resides roughly in the centre of the contour. To quantify our uncertainty, we compute the width and height of the rectangle circumscribing the contour. Then, we take the uncertainty in $\alpha_{c,\mathrm{exp}}$ to be half the width and the uncertainty in $\nu_{\mathrm{exp}}$ to be half the height.

This gives us the following experimental values for the critical point and critical exponent:
\begin{equation}
    \label{eq:CriticalPoints}
    \alpha_{c, \mathrm{exp}} = 0.95 \pm 0.06, \,\,\, \nu_{\mathrm{exp}} = 2.5 \pm 0.5.
\end{equation}

\section*{Data Availability}
The error-mitigated output from the quantum processors is available~\cite{zenodo_data} at the following Zenodo repository: \url{https://doi.org/10.5281/zenodo.7120441}. Data for Fig.~\ref{fig:experiment}, Fig.~\ref{fig:goodShotsProportion}, and Fig.~\ref{fig:Landscape} are also available in the repository.

\section*{Code Availability}
We used Qiskit~\cite{Qiskit} to execute the quantum circuits on the IBM Q quantum processors. All code used in classical and quantum simulation and analysis of experimental data is available~\cite{zenodo_data} at the following Zenodo repository: \url{https://doi.org/10.5281/zenodo.7120441}.

\section*{Acknowledgements}

This work was supported by the Minist\`{e}re de l'\'{E}conomie et de l'Innovation du Qu\'{e}bec via its contributions to its Research Chair in Quantum Computing and the IBM Q Hub of Institut quantique at Universit\'{e} de Sherbrooke. The work was also supported by a Natural Sciences and Engineering Research Council of Canada Discovery grant (S.K.), a B2X scholarship from the Fonds de recherche--Nature et technologies and a scholarship from the Natural Sciences and Engineering Research Council of Canada [funding reference number: 456431992] (J.C.). We acknowledge Calcul Qu\'{e}bec and Compute Canada for computing resources.

\section*{Author contributions}

J.C. conducted the experiments, performed all simulations and data analysis, made the figures, and wrote the paper. S.K. conceived the idea for the project, provided guidance along the way, and wrote the paper.

\section*{Competing interests statement}

We declare no competing interests.


\begin{thebibliography}{27}
\providecommand{\natexlab}[1]{#1}
\providecommand{\url}[1]{\texttt{#1}}
\expandafter\ifx\csname urlstyle\endcsname\relax
  \providecommand{\doi}[1]{doi: #1}\else
  \providecommand{\doi}{doi: \begingroup \urlstyle{rm}\Url}\fi

\bibitem[Côté and Kourtis(2022)]{cote_entanglement_2021}
C\^{o}t\'{e}, J. \& Kourtis, S.
\newblock Entanglement phase transition with spin glass criticality.
\newblock \emph{Phys. Rev. Lett.} \textbf{128,} 240601 (2022).
\newblock \url{https://doi.org/10.1103/PhysRevLett.128.240601}

\bibitem[Li et~al.(2019)Li, Chen, and Fisher]{li_measurement-driven_2019}
Li, Y., Chen, X. \& Fisher, M. P. A.
\newblock Measurement-driven entanglement transition in hybrid quantum
  circuits.
\newblock \emph{Phys. Rev. B} \textbf{100,} 134306 (2019).
\newblock \url{https://doi.org/10.1103/PhysRevB.100.134306}

\bibitem[Skinner et~al.(2019)Skinner, Ruhman, and
  Nahum]{skinner_measurement-induced_2019}
Skinner, B., Ruhman, J. \& Nahum, A.
\newblock Measurement-induced phase transitions in the dynamics of
  entanglement.
\newblock \emph{Phys. Rev. X} \textbf{9,} 031009 (2019).
\newblock \url{https://doi.org/10.1103/PhysRevX.9.031009}

\bibitem[Ippoliti et~al.(2021)Ippoliti, Gullans, Gopalakrishnan, Huse, and Khemani]{ippoliti_entanglement_2021}
Ippoliti, M., Gullans, M. J., Gopalakrishnan, S., Huse, D. A. \& Khemani, V.
\newblock Entanglement phase transitions in measurement-only dynamics.
\newblock \emph{Phys. Rev. X}, \textbf{11,} 011030, (2021).
\newblock \url{https://doi.org/10.1103/PhysRevX.11.011030}

\bibitem[Potter and Vasseur(2021)]{potter_entanglement_2021}
Potter, A. C. \& Vasseur, R.
\newblock Entanglement dynamics in hybrid quantum circuits.
\newblock In \emph{Entanglement in Spin Chains}. (eds Bayat, A., Bose, S. \& Johannesson, H.)
\newblock 211--249 (Springer, Cham, Switzerland, 2022).
\newblock \url{https://doi.org/10.1007/978-3-031-03998-0_9}

\bibitem[Noel et~al.(2021)]{noel_observation_2021}
Noel, C. et~al.
\newblock Measurement-induced quantum phases realized in a trapped-ion quantum computer.
\newblock \emph{Nat. Phys.} \textbf{18,} 760--764 (2022).
\newblock \url{https://doi.org/10.1038/s41567-022-01619-7}

\bibitem[Koh et~al.(2022)Koh, Sun, Motta, and Minnich]{koh_experimental_2022}
Koh, J. M., Sun S. N., Motta, M. \& Minnich, A. J.
\newblock Experimental realization of a measurement-induced
  entanglement phase transition on a superconducting quantum
  processor.
\newblock Preprint at
  \url{http://arxiv.org/abs/2203.04338} (2022).
  
\bibitem[Fisher et al(2022)]{fisher_random_2022}
Fisher, M. P.~A., Khemani, V., Nahum, A. \& Vijay, S.
\newblock Random quantum circuits.
\newblock Preprint at \url{http://arxiv.org/abs/2207.14280} (2022).
  
\bibitem[Lavasani et~al.(2021)Lavasani, Alavirad, and Barkeshli]{lavasani_topological_2021}
Lavasani, A., Alavirad, Y. \& Barkeshli, M.
\newblock Topological order and criticality in $(2+1)D$ monitored random quantum circuits.
\newblock \emph{Phys. Rev. Lett.} \textbf{127,} 235701, (2021).
\newblock \url{https://doi.org/10.1103/PhysRevLett.127.235701}

\bibitem[Lavasani et~al.(2022)Lavasani, Luo, and
  Vijay]{lavasani_monitored_2022}
Lavasani, A., Luo, Z.-X. \& Vijay, S.
\newblock Monitored quantum dynamics and the {Kitaev} spin liquid.
\newblock Preprint at \url{http://arxiv.org/abs/2207.02877} (2022).

\bibitem[Sriram et~al.(2022)Sriram, Rakovszky, Khemani, and
  Ippoliti]{sriram_topology_2022}
Sriram, A., Rakovszky, T., Khemani, V. \& Ippoliti, M.
\newblock Topology, criticality, and dynamically generated qubits in a stochastic measurement-only {Kitaev} model.
\newblock Preprint at \url{http://arxiv.org/abs/2207.07096} (2022).

\bibitem[Mézard et~al.(1984)Mézard, Parisi, Sourlas, Toulouse, and
  Virasoro]{mezard_nature_1984}
M\'{e}zard, M., Parisi, G., Sourlas, N., Toulouse, G. \& Virasoro, M.
\newblock Nature of the spin-glass phase.
\newblock \emph{Phys. Rev. Lett.} \textbf{52,} 1156 (1984).
\newblock \url{https://doi.org/10.1103/PhysRevLett.52.1156}

\bibitem[Schirber(2021)]{schirber_nobel_2021}
Schirber, M.
\newblock \emph{Physics}, \textbf{14,} 141 (2021).
\newblock \url{https://doi.org/10.1103/Physics.14.141}

\bibitem[Franz et~al.(2001)Franz, Mézard, Ricci-Tersenghi, Weigt, and
  Zecchina]{franz_ferromagnet_2001}
Franz, S., M\'{e}zard, M., Ricci-Tersenghi, F., Weigt, M. \& Zecchina, R.
\newblock A ferromagnet with a glass transition.
\newblock \emph{EPL (Europhy. Lett.)} \textbf{55,} 465 (2001).
\newblock \url{https://doi.org/10.1209/epl/i2001-00438-4}

\bibitem[Ricci-Tersenghi et~al.(2001)Ricci-Tersenghi, Weigt, and
  Zecchina]{ricci-tersenghi_simplest_2001}
Ricci-Tersenghi, F., Weigt, M. \& Zecchina, R.
\newblock Simplest random k-satisfiability problem.
\newblock \emph{Phys. Rev. E} \textbf{63,} 026702 (2001).
\newblock \url{https://doi.org/10.1103/PhysRevE.63.026702}

\bibitem[Sherrington and Kirkpatrick(1975)]{sherrington_solvable_1975}
Sherrington, D. \& Kirkpatrick, S.
\newblock Solvable model of a spin glass.
\newblock \emph{Phys. Rev. Lett.} \textbf{35,} 1792 (1975).
\newblock \url{https://doi.org/10.1103/PhysRevLett.35.1792}

\bibitem[Edwards and Anderson(1975)]{edwards_theory_1975}
Edwards, S. F. \& Anderson, P. W.
\newblock Theory of spin glasses.
\newblock \emph{J. Phys. F: Met. Phys.} \textbf{5,} 965 (1975).
\newblock \url{https://doi.org/10.1088/0305-4608/5/5/017}

\bibitem[Parisi(1983)]{parisi_order_1983}
Parisi, G.
\newblock Order parameter for spin-glasses.
\newblock \emph{Phys. Rev. Lett.} \textbf{50,} 1946 (1983).
\newblock \url{https://doi.org/10.1103/PhysRevLett.50.1946}

\bibitem[Sang and Hsieh(2021)]{sang_measurement-protected_2021}
Sang, S. \& Hsieh, T.~H.
\newblock Measurement-protected quantum phases.
\newblock \emph{Phys. Rev. Research}, \textbf{3} 023200 (2021).
\newblock URL \url{https://doi.org/10.1103/PhysRevResearch.3.023200}

\bibitem[Li and Fisher(2021)]{li_robust_2021}
Li., Y \& Fisher, M. P.~A.
\newblock Robust decoding in monitored dynamics of open quantum systems with $Z_2$ symmetry.
\newblock Preprint at \url{http://arxiv.org/abs/2108.04274} (2021).

\bibitem[Bao et~al.(2021)Bao, Choi, and Altman]{bao_symmetry_2021}
Bao, Y., Choi, S. \& Altman, E.
\newblock Symmetry enriched phases of quantum circuits.
\newblock \emph{Ann. of Phys.}, \textbf{435} 168618 (2021).
\newblock \url{https://doi.org/10.1016/j.aop.2021.168618}

\bibitem[Mézard et~al.(2003)Mézard, Ricci-Tersenghi, and
  Zecchina]{mezard_two_2003}
M\'{e}zard, M., Ricci-Tersenghi, F. \& Zecchina, R.
\newblock Two solutions to diluted p-spin models and {XORSAT} problems.
\newblock \emph{J. Stat. Phys.} \textbf{111,} 505--533 (2003).
\newblock \url{https://doi.org/10.1023/A:1022886412117}

\bibitem[Monasson(2007)]{monasson_course_2007}
Monasson, R.
\newblock In \emph{Complex {Systems}, {Les} {Houches}} Vol~85 (eds Bouchaud, J.-P., M\'{e}zard, M. \& Dalibard, J.) (Elsevier, 2007).
\newblock \url{https://doi.org/10.1016/S0924-8099(07)80008-4}

\bibitem[noa(2021)]{noauthor_ibm_2021}
{IBM} {Quantum}.
\newblock \url{https://quantum-computing.ibm.com/}, (2021).

\bibitem[Kawashima and Ito(1993)]{kawashima_critical_1993}
Kawashima, N. \& Ito, N.
\newblock Critical behavior of the three-dimensional ±{J} model in a magnetic
  field.
\newblock \emph{J. Phys. Soc. Jpn.} \textbf{62,} 435--438 (1993).
\newblock \url{https://doi.org/10.1143/JPSJ.62.435}

\bibitem[Cote and Kourtis(2022)]{zenodo_data}
C\^{o}t\'{e}, J. \& Kourtis, S.
\newblock Data for ``Qubit vitrification and entanglement criticality on a quantum processor''.
\newblock \emph{Zenodo} \url{https://doi.org/10.5281/zenodo.7120441} (2022).

\bibitem[et. al(2021)]{Qiskit}
ANIS, M. S. et. al.
\newblock Qiskit: An open-source framework for quantum computing, (2021).
\newblock \url{https://qiskit.org/}

\end{thebibliography}
\end{document}